\documentclass[preprint,showpacs,preprintnumbers,amsmath,amssymb]{revtex4}

\usepackage{graphicx}% Include figure files
\usepackage{dcolumn}% Align table columns on decimal point
\usepackage{bm}% bold math

\begin{document}

\preprint{APS/123-QED}

\title{Determination of the stellar (n,$\gamma$) cross section of $^{40}$Ca \\ with accelerator mass spectrometry}% Force line breaks with \\

\author{I. Dillmann}
\email{iris.dillmann@ph.tum.de}
 \altaffiliation[Present address: ]{Physik Department E12 and Excellence Cluster Universe, Tech\-nische Universit\"at M\"unchen, D-85748 Garching}
\author{C. Domingo-Pardo}
 \altaffiliation[Present address: ]{Gesellschaft f\"ur Schwerionenforschung, Darmstadt/ Germany}
\author{M. Heil}
 \altaffiliation[Present address: ]{Gesellschaft f\"ur Schwerionenforschung, Darmstadt/ Germany}
\author{F. K\"appeler}
 \affiliation{Forschungszentrum Karlsruhe, Institut f\"ur Kernphysik, Postfach 3640, D-76021 Karlsruhe}

\author{A. Wallner}
\author{O. Forstner}
\author{R. Golser}
\author{W. Kutschera}
\author{A. Priller}
\author{P. Steier}
 \affiliation{Vienna Environmental Research Accelerator, Fakult\"at f\"ur Physik, \\Universit\"at Wien, A-1090 Wien}
 
\author{A. Mengoni}
 \affiliation{International Atomic Energy Agency, Wagramer Strasse 5, A-1400 Wien}
 
\author{R. Gallino}
 \altaffiliation[also at ]{Centre for Stellar and Planetary Astrophysics, Monash University, Victoria 3800, Australia}
 \affiliation{Dipartimento di Fisica Generale, Universit{\`a} di Torino \\ Via P. Giuria 1, I-10125 Torino}

\author{M. Paul}
 \affiliation{Racah Institute of Physics, Hebrew University, IL-Jerusalem 91904}
 
\author{C. Vockenhuber} 
 \affiliation{TRIUMF, 4004 Wesbrook Mall, Vancouver, B.C., Canada V6T 2A3}

\date{\today}% It is always \today, today,

\begin{abstract}
The stellar $(n,\gamma)$ cross section of $^{40}$Ca at $kT$=25~keV has been measured with a combination of the activation technique and accelerator mass spectrometry (AMS). This combination is required in cases where direct off-line counting of the produced activity is compromised by the long half-life and/or missing $\gamma$-ray transitions. The neutron activations were performed at the Karlsruhe Van de Graaff accelerator using the quasi-stellar neutron spectrum of $kT$=25 keV produced by the $^7$Li($p,n$)$^7$Be reaction. The subsequent AMS measurements have been carried out at the Vienna Environmental Research Accelerator (VERA) with a 3 MV tandem accelerator. The doubly-magic $^{40}$Ca is a bottle-neck isotope in incomplete silicon burning, and its neutron capture cross section determines the amount of leakage and impacting thus on the eventual production of iron group elements. Due to its high abundance $^{40}$Ca can also play a secondary role as "neutron poison" for the $s$ process. Previous determinations of this value at stellar energies were based on time-of-flight measurements. Our method uses an independent approach, and yields for the Maxwellian-averaged cross section at $kT$=30~keV a value of $\langle\sigma\rangle_{30keV}$=5.73$\pm$0.34~mbarn.
\end{abstract}

\pacs{25.40.Lw, 26.30.-k, 27.40.+z, 97.10.Cv}% PACS, the Physics and Astronomy
\maketitle

\section{\label{intro}Introduction}
%\subsection{Stellar production}
The doubly-magic $^{40}$Ca is mainly produced in explosive burning processes during supernova explosions. In contrast to the ``normal'' hydrostatic burning phases the respective explosive burning is only ignited for a few seconds when the supernova shock front passes through the outer layers. The main fuels for the explosive burning stages are the so-called ``$\alpha$ nuclei'' consisting of $^{4}$He clusters ($^{12}$C, $^{16}$O, $^{20}$Ne, and $^{28}$Si). The ignition temperatures are between 1.9~GK (for carbon burning) up to 4~GK (for silicon burning).
The time scales of neon and silicon burning depend only on the respective temperatures and not on the density since both are dominated by the photodisintegration reactions of $^{20}$Ne$(\gamma,\alpha)$ and $^{28}$Si$(\gamma,\alpha)$.
Apart from its role in astrophysics, the ($n,\gamma$) cross section of $^{40}$Ca represents an important test for the applicability of the Hauser-Feshbach statistical model in the limit of low level densities. 

\subsection{Production of $^{40}$Ca}
Woosley \cite{woos73} showed already in 1973 that a superposition of explosive oxygen burning, complete silicon burning with $\alpha$-rich freeze-out, and incomplete silicon burning offers good fits to the solar abundances in the mass region 28$<A<$62. 

In explosive oxygen burning a quasi-statistical equilibrium (QSE) in the mass range $A$= 28--45 is obtained at $T$$>$3.3~GK \cite{MKc96,MKC98,FRR99}. This temperature is too low to achieve full nuclear statistical equilibrium (NSE) but in zones where the temperature exceeds 4~GK minor production of iron group elements can occur. The main burning products are the $\alpha$ nuclei $^{28}$Si, $^{32}$S, $^{36}$Ar, and $^{40}$Ca.

Explosive silicon burning occurs at $T$$>$4--5~GK and can be subdivided into incomplete burning, complete burning with normal freeze-out, and complete burning with $\alpha$-rich freeze-out. Complete Si burning is attained for $T$$>$5~GK where full NSE is established and the iron group elements are produced. Complete Si burning with $\alpha$-rich freeze-out occurs at low densities when the triple $\alpha$-reaction is not fast enough to keep the helium abundance in equilibrium. Then traces of $\alpha$ nuclei remain which were not transformed into iron group elements.

Incomplete silicon burning occurs at peak temperatures of $T$=4--5~GK when the temperature is not high enough for nuclear reactions to overcome the bottleneck at the proton-magic shell closure $Z$=20. The most abundant burning products are the same as for explosive oxygen burning, but partial leakage can produce iron group elements like $^{54}$Fe and $^{56}$Ni. Which of the three silicon burning scenarios actually takes place depends on the respective peak temperatures and densities during the passage of the supernova shock front.

\subsection{Neutron poisons in massive stars}
The neutrons for the $s$ process produced in situ by the $^{13}$C($\alpha,n$)$^{16}$O and $^{22}$Ne($\alpha,n$)$^{25}$Mg sources can be consumed by isotopes with high abundances $A_i$ and/or large neutron capture cross sections. The most important ``neutron poisons'' are $^{16}$O and $^{25}$Mg, the products of the neutron source reactions. These isotopes reduce the available neutrons per $^{56}$Fe seed and thus the $s$-process efficiency. Further neutron poisons are $^{17}$O, $^{20}$Ne, and $^{24}$Mg. This list can also be extended to $^{40}$Ca, since this doubly-magic isotope exhibits both, a rather high solar abundance which reflects the high abundance in the preceding generation of stars (see Table~\ref{tab:ca}), and one of the largest Maxwellian averaged cross sections among the 20 most abundant isotopes.

\begin{table}[!htb]
\caption{Isotopic \cite{iupac03} and solar abundances $A_i$ (relative to $A_{\rm{Si}}$= 10$^6$) \cite{ande89} of the stable Ca isotopes. \label{tab:ca}}
\renewcommand{\arraystretch}{1.2} % enlarge line spacing
\begin{ruledtabular}
\begin{tabular}{ccc}
Isotope		& Isotopic abundance [\%]  	& Solar abundance $A_i$\\
\hline
$^{40}$Ca	& 96.941 (156)							& 59200  \\
$^{42}$Ca	& 0.647 (23)								& 395 \\
$^{43}$Ca	& 0.135 (10)								& 82.5 \\
$^{44}$Ca	& 2.086 (110)								& 1275 \\
$^{46}$Ca	& 0.004	(3)									& 2.4 \\
$^{48}$Ca	& 0.187	(21)								& 114 \\
\hline
$\Sigma_{\rm{Ca}}$ & 100											& 61068.9 \\
\end{tabular}
\end{ruledtabular}
\end{table}

For the reasons outlined above, an accurate knowledge of the neutron capture cross section of $^{40}$Ca is of importance for the leakage through the $N$=$Z$=20 bottleneck and for the role as possible neutron poison. The previous recommended Maxwellian-averaged cross section (MACS) of $\langle \sigma \rangle_{30keV}$=6.7$\pm$0.7~mb at $kT$=30~keV \cite{bao00} was taken directly from de L. Musgrove et al. \cite{musg76b} and is essentially based on the respective resonance parameters measured below $E_n$=300~keV, with contributions from the total neutron cross section measurement of Singh et al. \cite{SLR74}, both using the time-of-flight (TOF) method. 

The remeasurement of the $^{40}$Ca$(n,\gamma)$$^{41}$Ca cross section with an independent method such as the activation technique plus following AMS could add confidence in these results. The very high sensitivity of the activation technique allows one to use very small samples so that neutron scattering corrections are completely negligible. A further advantage of the activation technique compared to the TOF method is that the direct capture component (DRC) is already included. In combination with accelerator mass spectrometry (AMS) this technique can be extended to hitherto inaccessible cases, where the product nuclei exhibit a long half-life and no $\gamma$-ray transitions (e.g. $^{41}$Ca, $t_{1/2}$= 103000 a \cite{PAK91,PAK92}). Direct atom counting via AMS also avoids uncertainties associated with the $\gamma$-decay branching of the activation products.

In Sec.~\ref{exp} we describe the experimental techniques of the activation with a stellar neutron source and the subsequent counting of the produced $^{41}$Ca with accelerator mass spectrometry. Sec.~\ref{data} shows the data analysis, and the results are presented in Sec.~\ref{res}. The astrophysical conclusions of our results are given in Sec.~\ref{astro}.

\section{Experimental technique}\label{exp}
\subsection{Activation}
The activation was carried out at the Karlsruhe 3.7 MV Van de
Graaff accelerator. Neutrons were produced with the
$^7$Li($p,n$)$^7$Be source by bombarding 30 $\mu$m thick layers of
metallic Li on a water-cooled Cu backing with protons of 1912 keV,
30 keV above the reaction threshold. The angle-integrated neutron
spectrum imitates almost perfectly a Maxwell-Boltzmann
distribution for $kT$ = 25.0$\pm$0.5~keV with a maximum neutron
energy of 106~keV \cite{raty88}. 
%Hence, the proper stellar capture
%cross section can be directly deduced from our measurement. 
At this proton energy the
neutrons are kinematically collimated in a forward cone with
120$^\circ$ opening angle. Neutron scattering through the Cu
backing is negligible since the transmission is about 98 \% in the
energy range of interest. To ensure homogeneous illumination of
the entire surface, the proton beam with a DC current of
$\approx$100~$\mu$A was wobbled across the Li target. Thus, the
mean neutron intensity over the period of the activations was
$\approx$1.0--1.6$\times$10$^9$ s$^{-1}$ at the position of the
samples, which were placed in close geometry to the Li target. A
$^6$Li-glass monitor at 1~m distance from the neutron target was used to
record the time-dependence of the neutron yield in intervals of
1~min as the Li target degrades during the irradiation. In this
way the proper correction of the number of nuclei which
decayed during the activation can be attained. This correction is negligible for very
long half-lives, e.g. for $^{41}$Ca, but becomes important for comparably
short-lived isotopes and affects the determination of the $^{198}$Au activity, since gold is used as the reference cross section for the neutron fluence determination.

A sample of calcium carbonate (152.6~mg CaCO$_3$, corresponding to 8.926$\times$10$^{20}$ atoms $^{40}$Ca) with natural isotopic abundance (96.941\% $^{40}$Ca \cite{iupac}, see Table~\ref{tab:ca}) was used in this measurement. The
sample material was pressed into a thin pellet 8~mm in diameter,
enclosed in 15~$\mu$m thick aluminium foil and
sandwiched between two 30~$\mu$m thick gold foils of the same
diameter. In this way the neutron fluence in our experimental neutron distribution 
can be determined relative to the well-known capture cross section of $^{197}$Au \cite{raty88}. During the activation the gold foils and the Li targets were changed three times, subdividing the experiment into four single activations (see Table~\ref{tab:act}). The net irradiation time was $\approx$20~d, and the total neutron exposure was calculated using the activated gold foils to be 2.32$\times$10$^{15}$ neutrons.

\begin{table}[!htb]
\caption{Overview of the activation parameters. $"t_a"$ is the
irradiation time and ``$\Phi_{tot}$'' the total neutron exposure. For discussion of the uncertainties see Sec.~\ref{error}.\label{tab:act}}
\renewcommand{\arraystretch}{1.2} % enlarge line spacing
\begin{ruledtabular}
\begin{tabular}{cccc}
Activation & $t_a$ [min] & \multicolumn{2}{c}{$\Phi_{tot}$ [$\times$10$^{14}$n]} \\
\hline 
ca-a & 4075  & \multicolumn{2}{c}{2.542} \\
ca-b & 10078 & \multicolumn{2}{c}{7.085} \\
ca-c & 9909  & \multicolumn{2}{c}{9.606} \\
ca-d & 4123  & \multicolumn{2}{c}{4.008} \\
\hline
Total & 28185 & \multicolumn{2}{c}{23.242} \\
\end{tabular}
\end{ruledtabular}
\end{table}

\subsection{AMS measurement}

The number of $^{41}$Ca nuclei produced in the neutron activation was measured offline applying the technique of accelerator mass spectrometry (AMS) at the VERA (Vienna Environmental Research Accelerator) facility. This facility is based on a 3-MV tandem accelerator equipped with a 40-sample caesium sputter source. Details on Ca measurements at VERA can be found in Refs. \cite{WGK06,WDG07}. The sample material was sputtered in the ion source and the nuclides are analyzed according to their mass and energy. Calcium fluoride (CaF$_2$) was chosen as proper sputter material. It was shown to provide sufficiently large negative currents when extracting CaF$_{3}$$^{-}$ ions, with the additional advantage that the production of the interfering isobar (KF$_{3}$$^-$) is strongly reduced \cite{UKN86, KEC89, FMK90}.

After the neutron activation, the entire Ca-containing pellet (CaCO$_{3}$) was dissolved in nitric acid to ensure a uniform mixture of the irradiated sample material and, consequently,  a cross-section value integrated over the full sample area is obtained. Several aliquots were used to prepare sputter targets by the following way: Under pH control, hydrofluoric acid (HF) was added to the solution and Ca precipitated as CaF$_{2}$, separated via centrifugation, and then dried to get CaF$_{2}$ powder. After mixing with Cu powder this material was pressed into the sample holders and inserted into the ion source. Overall, 17 sputter cathodes and a similar amount of identical but non-irradiated blank samples were prepared for the AMS measurements. 

The $^{41}$Ca content was determined during eight measurement series applying two different particle detection setups reducing thus systematic uncertainties. AMS measures isotope ratios by determining count rate of the rare isotope ($^{41}$Ca) relative to the ion current of a reference isotope ($^{40}$Ca). The procedure applied at VERA is the following: Negatively charged ions are mass separated and injected into a tandem accelerator which runs at typically 3 MV. At the terminal the particles have to pass a gas stripper canal where electrons are stripped off from the negative ions and parasitic molecular ions dissociated via collisions and Coulomb break-ups in the gas. The resulting positive ions are further accelerated. After the tandem accelerator the 4$^{+}$ charge state is selected: i.e. particles having energies of 13.5 MeV and the right mass over charge ratio ($M/q$) are further transported to the particle detection setup. The number of radionuclides is measured with an energy-sensitive detector, while the current of the stable ions is determined with Faraday cups. In case of Ca, fast switching between stable isotopes and the radionuclide was applied at the injector magnet and both, $^{40}$Ca and $^{42}$Ca currents, were registered. The isotope ratios of the unknown samples were compared with ratios of $^{41}$Ca/$^{40}$Ca reference materials \cite{NCD00}. Blank samples, having negligible $^{41}$Ca concentrations, were used as controls for the quantification of the background level.

Two different detection setups were used for these measurements, both with the goal to separate efficiently possible interferences in the signal of the radioisotope $^{41}$Ca with the isobar $^{41}$K. The first uses the so-called delta~TOF-setup \cite{VGK05,SGK05}. Here, the time-of-flight is measured for particles which pass a thick but very homogeneous, passive absorber foil made of silicon nitride (Silson Ltd., UK). K and Ca can be discriminated due to their different energy loss in the absorber, which is reflected in a different velocity when leaving the foil. The second method utilizes a recent development at ETH Zurich/Paul Scherrer Institut (PSI) of a compact type of ionization chamber which provides a high resolution in the energy measurement \cite{SJS00}. Particles enter this detector through a sufficiently thick silicon nitride entrance foil, and isobars, dispersed in energy by the loss in the foil, are discriminated with a segmented anode via a $\Delta$$E-E$ measurement. 

For the measurement of the (n,$\gamma$) cross section we expected an isotope ratio $^{41}$Ca/Ca of $\approx$10$^{-11}$, which is well above the machine background of $^{41}$Ca/Ca at VERA and the level of blank and calcium material used as target for neutron activation. Both setups have been proven to provide a sufficient isobar suppression for our samples.

\section{Data analysis and results}\label{data}
\subsection{Determination of the neutron flux}
The measurement of the induced $^{198}$Au activity after the
irradiation was performed with a high
purity germanium (HPGe) detector with a well defined measuring position at a distance
of 76~mm surrounded by 10~cm lead shielding. The absolute efficiency for the
411.8~keV $\gamma$-line of the decay into $^{198}$Hg was
determined with a set of reference sources and yielded $\varepsilon_\gamma$=0.212\%.

The total amount of activated $^{198}$Au nuclei, $N_{198}$, at the end of the
irradiation can be deduced from the number of events $C(t_m)$ in the
particular $\gamma$-ray line at 411.8~keV registered in the HPGe detector
during the measuring time $t_m$ \cite{beer80}. The factor
$t_w$ corresponds to the waiting time between the end of
the irradiation and the start of the activity measurement:
\begin{eqnarray}
N_{198} = \frac{C(t_m)} {\varepsilon_\gamma~ I_\gamma~
k_{\gamma}~(1-e^{-\lambda~t_m})~e^{-\lambda~t_w}} \label{Z}.
\end{eqnarray}
$I_\gamma$ accounts for the relative $\gamma$ intensity per decay
of the 411.8~keV transition ($I_\gamma$= 95.58$\pm$0.12 \% \cite{NDS198}). For the measurement of the activated gold foils with the HPGe, the $\gamma$-ray self-absorption $k_\gamma$ has to be
considered \cite{beer80}. This correction factor was 0.995 for all gold foils. 

The number of activated sample atoms $N_{act}$ is determined by
\begin{eqnarray}
N_{act}=N_i~\langle\sigma_{exp}\rangle(i)~\Phi_{tot}~f_b(i). \label{act}
\end{eqnarray}
In this equation, $\langle\sigma_{exp}\rangle$(i) is the cross section in the
experimental neutron spectrum and $\Phi_{tot}$ the total neutron
flux. The factor
\begin{eqnarray}
f_b=\frac{\int_{0}^{t_a}\phi(t)~e^{-\lambda(t_a-t)}~dt}{\int_{0}^{t_a}\phi(t)~dt}
\label{fb}
\end{eqnarray}
accounts for the decay of activated nuclei during the irradiation
time $t_a$ as well as for variations in the neutron flux.
This factor is calculated from the neutron flux history recorded
throughout the irradiation with the $^6$Li glass detector at 1~m distance from the target. The reference value for the experimental $^{197}$Au cross section in the quasi-stellar spectrum of the
$^{7}$Li$(p,n)$$^{7}$Be source is $\langle\sigma_{exp}\rangle$($^{197}$Au)=586$\pm$8~mbarn \cite{raty88}.

The time-integrated neutron flux $\Phi_{tot} = \int \phi(t)dt$ seen by the sample (Table~\ref{tab:act}) is determined by averaging the neutron fluxes of the two gold foils enclosing the
respective sample:
\begin{eqnarray}
\Phi_{tot}= \frac{N_{198}}{N_{197}~\langle\sigma_{exp}\rangle(^{197}\rm{Au})~f_b}. \label{act2}
\end{eqnarray}

\subsection{Determination of the isotopic ratio via AMS}
Accelerator mass spectrometry determines the ratio of a radioactive isotope relative to one stable isotope of the same element. This measurement is carried out relative to reference standards with well-known isotopic ratios. In our case we have determined the ratio of $^{41}$Ca versus $^{40}$Ca, $R$=$\frac{N_{41}}{N_{40}}$. 

\begin{figure}[!htb]
\includegraphics[scale=1]{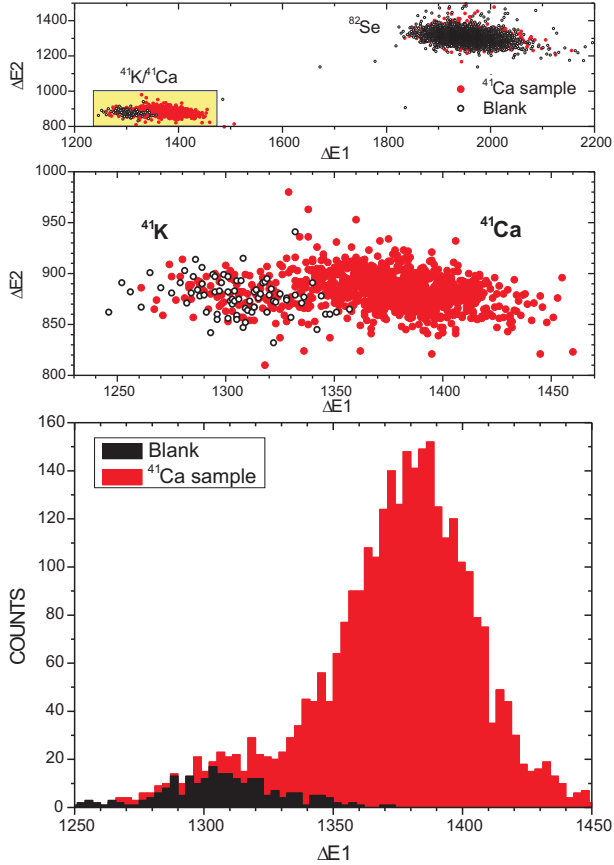}% Here is how to import EPS art
\caption{\label{ams1} (Color online) Top/Middle: First ($\Delta$E1) and second ($\Delta$E2) energy loss signal from the anodes in the ionization chamber. The events from $^{41}$Ca and its stable isobar $^{41}$K can be distinguished by comparing a blank and an irradiated sample with $^{41}$Ca. In the top panel events from $^{82}$Se appear at higher energy losses which have the same $M/q$ as $^{41}$Ca. Bottom: Histogram illustrating the separation between $^{41}$K and $^{41}$Ca.}
\end{figure}

Fig.~\ref{ams1} shows the first and second energy loss signal from the ionization chamber measured with a blank sample and an irradiated sample with $^{41}$Ca. The events from the quasi-stable isobar $^{41}$K and the radioisotope $^{41}$Ca can be clearly distinguished. With these settings, also ions with the same $M/q$ as $^{41}$Ca can reach the detector, like $^{82}$Se. However, as can be seen in the top part of Fig.~\ref{ams1}, these events are well separated from the $A$=41 signals.

The AMS measurements were repeated eight times. In the first six measurements the deltaTOF technique was applied, and in the remaining two the compact ionization chamber (see section II.B). The isotope ratio $R$=$\frac{N_{41}}{N_{40}}$ from these measurements is (1.34$\pm$0.07)$\times$10$^{-11}$. The uncertainty is due to the 5.2\% error (1$\sigma$) from the AMS measurement. 

One can determine the experimental cross section $\langle\sigma_{exp}\rangle$ for the $^{40}$Ca($n,\gamma$) reaction for our experimental neutron spectrum with the known total neutron exposure $\Phi_{tot}$= 2.324$\times$10$^{15}$ from Table~\ref{tab:act} and Eq.~\ref{act}:
\begin{eqnarray}
\langle\sigma_{exp}\rangle(^{40}\rm{Ca})=  \frac{N_{41}}{N_{40}}~\frac{1}{\Phi_{tot} \cdot f_b}. \label{act3}
\end{eqnarray}
Since $\frac{N_{41}}{N_{40}}$ is the isotopic ratio $R$, and the correction factor $f_b$ for very long half-lives is equal to unity, this equation simplifies to
\begin{eqnarray}
\langle\sigma_{exp}\rangle(^{40}\rm{Ca})=  R~\frac{1}{\Phi_{tot}}. \label{act4}
\end{eqnarray}
From this equation we can deduce an experimental cross section of $\langle\sigma_{exp}\rangle(^{40}\rm{Ca})$= 5.77$\pm$0.34~mbarn (uncertainty including the total uncertainty from Table~\ref{tab:err}).

\subsection{\label{err}Error analysis}\label{error}
The experimental uncertainties are summarized in Table~\ref{tab:err}. Since stellar neutron capture cross section measurements are often carried out relative to gold \cite{bao00}, the error of 1.4\% \cite{raty88} in the gold cross section cancels out in most astrophysical applications and was, therefore, not included in the present error analysis.
An error of 2.0\% was assumed to account for the uncertainty of the sample position (0.25~mm) relative to the gold foils during the activation that affects the neutron flux seen by the sample. The uncertainty of the HPGe efficiency calibration was derived from the accuracy of the calibration sources and of the calibration procedure. The uncertainty (1~$\sigma$) in the AMS measurements (5.2\%) includes the statistical uncertainty, the reproducibility of the measurement and a systematic contribution from the measurement relative to reference materials (which is dominated by the uncertainty of the half-life of $^{41}$Ca). The total uncertainty of 5.9\% is the combination of the uncertainties from the neutron flux measurement (2.9\%) and the AMS measurement (5.2\%).

\begin{table}[!htb]
\caption{Uncertainties for $^{197}$Au and $^{40}$Ca. \label{tab:err}}
\renewcommand{\arraystretch}{1.2} % enlarge line spacing
\begin{ruledtabular}
\begin{tabular}{ccc}
		& \multicolumn{2}{c}{Uncertainty (\%)} \\
		 	Source of uncertainty 	& $^{197}$Au & $^{40}$Ca \\
\hline
		Gold cross section & 1.4\footnotemark[1] & -- \\
		Detector efficiency & 2.0 & -- \\
		Neutron flux & 2.0 & -- \\
		Sample mass & 0.4 & -- \\
		$\gamma$-ray intensity & 0.1 & -- \\
		Counting statistics & 0.2 & -- \\
	  AMS uncertainty & -- & 5.2 \\
\hline
	  Total uncertainty & 2.9 & 5.9\footnotemark[2] \\
\end{tabular}
\end{ruledtabular}
\footnotetext[1]{Not included in the final uncertainty, see text.}
\footnotetext[2]{Including the 2.9\% uncertainty from the flux measurement with the gold samples.}

\end{table}

\section{Cross sections}\label{res}
\subsection{Evaluated cross section libraries}
The energy-dependent neutron cross sections from the evaluated cross section libraries JEFF-3.0A, JEFF-3.1, ENDF/B-VII.0, and JENDL-3.3 include experimental resonance parameters adopted from the compilations of Mughabghab \cite{mugh81,mugh06}. These parameters were determined in a TOF measurement by de L. Musgrove \cite{musg76b} in the energy range between 2.5 and 300~keV with contributions from the total neutron cross section measurement of Singh et al. \cite{SLR74}. The quoted uncertainty for most resonance parameters from de L. Musgrove \cite{musg76b} is of the order of 10\% and larger. 

The unresolved resonance region above 296~keV is based on different statistical model calculations and thus differs from evaluation to evaluation. For example, in JENDL-3.3 this region originates from the code CASTHY \cite{casthy} and was normalized to the Maxwellian average $\langle\sigma\rangle_{30keV}$= 6.7~mbarn given in \cite{bao00}. The data sets of JEFF-3.1 and ENDF/B-VII.0 are identical and use the recent nuclear model code TALYS \cite{talys}, where the parameters have been adjusted to reproduce the existing experimental data. However, for the calculation of Maxwellian cross sections up to $kT$=100~keV the contribution from the unresolved resonance region has only little influence and we used an average of all four data sets.

\subsection{Maxwellian averaged cross sections}
In an astrophysical environment with temperature $T$ interacting 
particles are quickly thermalized by collisions in the stellar 
plasma, and the neutron energy distribution can be described by a Maxwell-Boltzmann spectrum:
\begin{eqnarray}
\Phi = dN/dE_n \sim \sqrt{E_n} \cdot e^{-E_n /kT} \label{eq:phi}.
\end{eqnarray}                 
The experimental neutron spectrum of the $^7$Li($p,n$)$^7$Be
reaction simulates the energy dependence of the flux $v \cdot \Phi \sim E_n \cdot e^{-E_n /kT}$ with
$kT$=25.0 $\pm$ 0.5~keV almost perfectly \cite{raty88}. However, 
the cutoff at $E_n$= 106~keV and small deviations from the 
shape of the ideal Maxwellian spectrum require a correction 
of the measured cross section $\langle\sigma_{exp}\rangle$ for obtaining a 
true Maxwellian average, $\langle\sigma\rangle$$_{25~keV}$. This correction
is determined by means of the energy-dependent cross 
sections from data libraries.

Before calculating a Maxwellian averaged cross section, we determine a "normalization factor" $NF$ which gives a direct comparison how good the agreement is between our measured cross sections $\langle\sigma_{exp}\rangle$ and the evaluated cross sections $\sigma(E)$ folded with the experimental neutron spectrum of the $^7$Li$(p,n)$$^7$Be source \cite{raty88}, $\langle\sigma_{eval}\rangle$=5.18~mbarn. Our experimentally determined cross section is $\langle\sigma_{exp}\rangle$=5.77~mbarn, resulting in a normalization factor of $NF$=$\frac{\langle\sigma_{exp}\rangle}{\langle\sigma_{eval}\rangle}$=1.11. 

The respective cross section derived by folding the NON-SMOKER energy dependence \cite{rau00,nons} with our experimental neutron distribution would be $\langle\sigma_{\rm{NS}}\rangle$=12.48~mbarn, yielding a normalization factor of $NF$=0.46. The Hauser-Feshbach model cannot be applied below a given minimum energy of $kT$$\approx$40~keV ($T_9\approx$0.48~GK) \cite{nons} due to the low level density of the double-magic nucleus $^{40}$Ca which causes an overestimation of cross sections. Such low level densities occur also for very neutron- or proton-rich isotopes close to the driplines, low mass nuclei, and at low energies. 

Applying the normalization factor $NF$ to the evaluated libraries affects 
also the corresponding thermal value and the strength of the individual 
resonances. However, since the uncertainty of the resonance parameters is, as mentioned above, 10\% and larger, we decided to use this approach. As soon as more accurate resonance parameters become available for $^{40}$Ca, this data should supersede our way of extrapolating Maxwellian cross sections via Eq.~\ref{eq:macs}:
\begin{eqnarray}
\langle\sigma\rangle_{kT}=\frac{2}{\sqrt{\pi}}~\cdot \frac{\int \frac{\sigma(E_n)}{NF}~E_n~e^{-E_n /kT} ~dE_n}{\int
E_n~e^{-E_n /kT}~dE_n} ~\label{eq:macs}.
\end{eqnarray}
The factor 2/$\sqrt{\pi}$ comes from the normalization of the Maxwellian flux formula when using the most probable velocity instead of the mean thermal velocity. At $kT$=30~keV this yields $\langle\sigma\rangle_{30keV}$=5.73$\pm$0.34~mbarn with our experimentally derived normalization factor $NF$=1.11.

We have also re-calculated the Maxwellian averaged cross sections using the resonance parameters given in ENDF/B-VII.0 \cite{endfb7} with the R-matrix analysis code SAMMY \cite{sammy}, yielding $\langle \sigma \rangle_{30keV}$= 5.13~mbarn at $kT$=30~keV. A calculation of the uncertainties was done independently by  Monte Carlo sampling of the four most important resonance parameters which contribute with 85.3\% to the Maxwellian average at $kT$=30~keV. Varying the resonances at E$_n$=10.83, 20.43, 42.12, and 52.57~keV within their given uncertainties yielded $\langle\sigma\rangle_{30~keV}$=5.04$\pm$0.30~mbarn. Varying the neutron widths $\Gamma_n$ of all resonances by 5\% and the respective Gamma widths $\Gamma_\gamma$ by 20\% gives $\langle\sigma\rangle_{30~keV}$=5.16$\pm$0.31~mbarn, in perfect agreement with the SAMMY calculation.

In Table~\ref{tab:macs} these values are compared to the previously recommended values from Bao et al. \cite{bao00}, which were adopted from Ref.~\cite{BVW92} and originate directly from Ref. \cite{musg76b}. As can be seen in Table~\ref{tab:macs} and also in the graphical comparison of the energy trends in Fig.~\ref{macs2}, the differences between the values from \cite{musg76b,BVW92,bao00} and the recalculated Maxwellian cross sections increase with increasing temperature.

\begin{table}[!htb]
\caption{\label{tab:macs}Maxwellian averaged cross sections
$\langle\sigma\rangle_{kT}$ (in mbarn) for $^{40}$Ca calculated with the energy dependencies from evaluated libraries ("This work"). The second column shows the previous recommended values \cite{bao00,BVW92} based on the value from \cite{musg76b}, in the third column the results calculated with resonance parameters in ENDF/B-VII.0 \cite{endfb7} are given.}
\begin{ruledtabular}
\renewcommand{\arraystretch}{1.1} % enlarge line spacing
\begin{tabular}{cccc}
$kT$    & \multicolumn{3}{c}{$\langle\sigma\rangle_{kT}$ [mbarn]} \\
$$[keV] & This work & Ref. \cite{musg76b,bao00,BVW92} & Ref. \cite{endfb7} \\
\hline
5  & 11.96 & 11.3 & 10.79 \\ 
10 & 11.49 & 12.1 & 10.35 \\
15 & 9.23 & 10.3 & 8.31 \\
20 & 7.56 & 8.6 & 6.81 \\
25 & 6.46 & 7.5 & 5.81 \\
30 & 5.73 (34) & 6.7 (7) & 5.16 (31)\\
40 & 4.93 & 5.8 & 4.44 \\
50 & 4.53 & 5.4 & 4.08 \\
60 & 4.28 & 5.2 & 3.85 \\
80 & 3.91 & 4.9 & 3.51 \\
100 & 3.59 & 4.9 & 3.19 \\
\end{tabular}
\end{ruledtabular}
\end{table}

%Furthermore Maxwellian averaged cross sections have to be corrected by a temperature-dependent stellar enhancement factor SEF(T)=$\frac{\sigma^{star}}{\sigma^{lab}}$. The stellar cross section $\sigma^{star}$=$\sum_{\mu} \sum_{\nu} \sigma^{\mu\nu}$ accounts for all transitions from excited target states $\mu$ to final states $\nu$ in thermally equilibrated nuclei, whereas the laboratory cross section $\sigma^{lab}$=$\sum_{0} \sum_{\nu} \sigma^{0\nu}$ includes only captures from the target ground state. These factors are tabulated, e.g. in Refs.~\cite{rau00,nons,bao00}. The SEF for $^{40}$Ca is equal to unity in the entire temperature  range up to $kT$=100~keV.

\subsection{Comparison with previous values}
Table~\ref{tab:comp} compares in the upper part the extrapolated Maxwellian averaged cross sections at $kT$=30~keV from the evaluated library ENDF/B-VII.0 \cite{endfb7} (using data from de L. Musgrove \cite{musg76b} and Singh \cite{SLR74}), Barrett \cite{barr76}, and the Bao et al. compilation \cite{bao00,BVW92} (taken directly from \cite{musg76b}) with our activation cross section. In the lower part of Table~\ref{tab:comp} the results from four Hauser-Feshbach predictions are given. Incidentally, the value of Woosley \cite{woos78} agrees best with our experimental value of 5.73$\pm$0.34 mbarn, but also the recalculated value from the resonance parameters ($\langle\sigma\rangle_{30keV}$=5.16$\pm$0.31~mbarn) agrees within the uncertainties. The large discrepancies to recent NON-SMOKER \cite{rau00,nons} and MOST \cite{most02,most05} results indicate in the present case the limited reliability of the statistical model in view of the low level density in the double-magic nucleus $^{40}$Ca in this energy region.

\begin{table}[!htb]
\caption{Comparison of Maxwellian averaged cross sections at $kT$=30~keV from previous experiments and theoretical predictions. Theoretical values are given without error bars.\label{tab:comp}}
\renewcommand{\arraystretch}{1.2} % enlarge line spacing
\begin{ruledtabular}
\begin{tabular}{lcc}
Reference & $\langle\sigma\rangle_{30keV}$  			& Ratio to \\
					&		[mbarn]															& this work \\
\hline 
This work  											& 5.73 (34) & -- \\
ENDF/B-VII.0 \cite{endfb7}			& 5.16 (31)	& 0.90$^{+0.11}_{-0.10}$ \\
Barrett \cite{barr76} 					& 10 (3) 		& 1.75$^{+0.67}_{-0.59}$ \\
Bao \textit{et al.} \cite{musg76b,bao00,BVW92} & 6.7 (7)		& 1.17$^{+0.20}_{-0.18}$ \\
\hline
Woosley \cite{woos78} 			& 5.8				& 1.01$^{+0.06}_{-0.06}$ \\
NON-SMOKER \cite{rau00} 		& 13.0			& 2.27$^{+0.14}_{-0.13}$ \\
MOST (2002)	\cite{most02} 	& 24.1			& 4.21$^{+0.27}_{-0.24}$ \\
MOST (2005)	\cite{most05} 	& 13.9			& 2.43$^{+0.15}_{-0.14}$ \\
\end{tabular}
\end{ruledtabular}
\end{table}

\subsection{Statistical model and direct capture calculations}
Despite the fact that the Hauser-Feshbach model cannot (or only with large uncertainties) be applied below $kT$$\approx$40~keV we have performed statistical model calculations using the code "HFSM" with standard optical model parameters from Moldauer \cite{mold63, mold64}. The level density is based on the Fermi gas theory with pairing and shell corrections. The parameters were fitted to reproduce the average level spacings of Mughabghab \cite{mugh81}, e.g. for $s$ waves $\langle$D($l$=0)$\rangle$=45$\pm$4~keV. The giant dipole resonance parameters for the calculation of the average radiative widths were $E$=20.14~MeV with a width of 8.05~MeV and a peak cross section of 58.3~mbarn. These values yield an average radiative width of $\langle\Gamma_\gamma\rangle_s$=193 meV, which is a factor of $\approx$8 lower than the 1.5$\pm$0.9~eV from de L. Musgrove \cite{musg76b}. However, only with this value reasonable Maxwellian averaged cross sections can be obtained. This decrease is interesting since the large difference of almost one order of magnitude to the average $p$- and $d$-wave widths ($\langle\Gamma_\gamma\rangle_p$=360$\pm$90~meV and $\langle\Gamma_\gamma\rangle_d$=740$\pm$360~meV) discussed in Ref. \cite{musg76b} vanishes.

Additional calculations were performed with the data set of ENDF/B-VII.0 \cite{endfb7}, based on the evaluation of Mughabghab \cite{mugh81,mugh06}, and with the model code TALYS \cite{talys} for cross-checking. The results are consistent within a few percent for energies up to 10~keV, but differ by up to 20\% for higher energies. TALYS calculations were repeated using a coupled channel code but this did not influence the neutron capture cross section. Thus the different energy dependencies from ENDF/B-VII.0 and TALYS originated only from the different optical model potentials used.

Direct radiative capture (DRC) calculations have been performed for $s$- and $p$-wave neutron transitions to several bound states in $^{41}$Ca. Using the experimental thermal cross section of $\sigma_{th}$=410$\pm$20~mbarn \cite{mugh06} and subtracting the contribution of 13.5~mbarn from the tail of the $s$-wave resonances the estimated DRC cross section is 396.5~mbarn. The extrapolation from thermal energies to the keV region leads to very small DRC contributions of 0.4~mbarn and 5.4~$\mu$barn for $s$- and $p$-wave neutrons at $kT$=25~keV, respectively. 

%For the next doubly-magic isotope in the Ca chain, $^{48}$Ca, the situation is completely different. Theoretical calculations \cite{KBO96} in comparison with experimental results at $kT$=25~keV \cite{KWM85} showed that the neutron capture cross section for this nucleus is completely dominated by the DRC component due to the absence of any significant resonances.

\begin{figure}[!htb]
\includegraphics[scale=1.05]{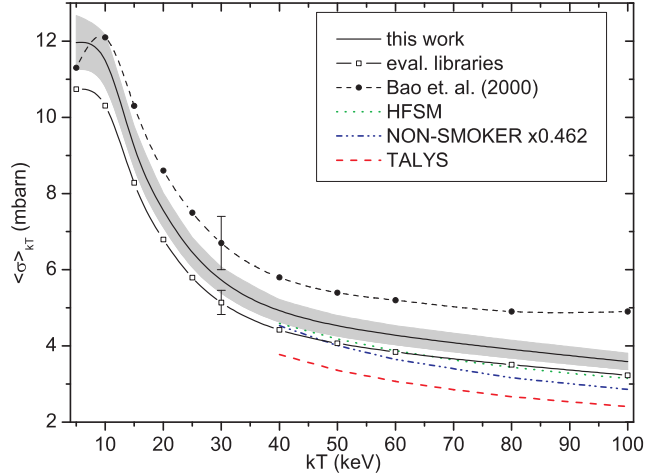}% Here is how to import EPS art
\caption{\label{macs2}(Color Online) Maxwellian averaged cross sections from this work compared with previously recommended values \cite{musg76b,bao00,BVW92} and the values calculated from resonance parameters of ENDF/B-VII.0 \cite{endfb7} (Table~\ref{tab:macs}). The error band corresponds to the 5.9\% uncertainty of our present measurement. Also plotted are the results obtained from the HF calculations HFSM and Talys, and the normalized NON-SMOKER predictions.}
\end{figure}

The best estimate for the Maxwellian averaged cross sections in the keV region can thus be obtained directly from the resonance parameters given in \cite{mugh81,mugh06} normalized to the present results. For neutron energies below $E_n$=100~keV, Maxwellian cross sections are determined by single resonances, and the Hauser-Feshbach theory cannot be applied. The predicted Maxwellian averaged cross sections in Fig.~\ref{macs2} should therefore be taken with care and are only plotted for $kT$$>$40 keV.

\begin{figure}[!htb]
\includegraphics[scale=1.05]{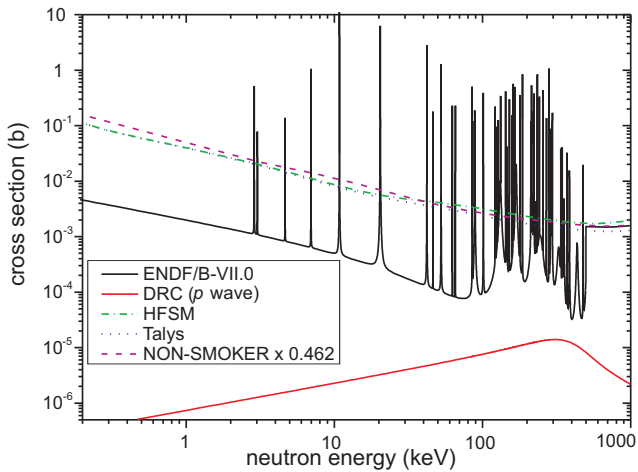}% Here is how to import EPS art
\caption{\label{dxs}(Color online) Energy-dependent cross sections $\sigma(E)$ for $^{40}$Ca from the evaluated data library ENDF/B-VII.0 (black line) in comparison with the prediction from the Hauser-Feshbach models HFSM, NON-SMOKER (normalized by a factor of 0.462 to reproduce the experimental cross section), and TALYS. The red line in the lower part shows the DRC contribution from $p$ wave neutrons.}
\end{figure}

\section{Conclusions}\label{astro} 
We have measured the $^{40}$Ca$(n,\gamma)$ cross section via the activation technique in a stellar neutron distribution and derived a Maxwellian averaged cross section of $\langle\sigma\rangle_{30keV}$=5.73$\pm$0.34~mbarn at $kT$=30~keV. This value is, within the error bars, in good agreement with the previously recommended value of 6.7$\pm$0.7~mbarn \cite{bao00,BVW92} taken directly from Ref.~\cite{musg76b}, and higher than the value calculated from the resonance parameters \cite{endfb7}, $\langle\sigma\rangle_{30keV}$=5.16$\pm$0.31~mbarn. With this result, $^{40}$Ca still acts as a significant neutron poison in stellar environments, simply due to its high initial abundance. In the direct $s$-process flow $^{40}$Ca is bypassed by its quasi-stable isobar $^{40}$K. In thermally pulsing low mass AGB stars \cite{gall98} $^{40}$Ca is therefore depleted by 13\% in the $s$ process. However, the further reaction flow via $^{41}$Ca can reach equilibrium. Accordingly, ($n,\gamma$) reactions on $^{40}$Ca contribute to the overall $s$-process efficiency in the Ca to Cr region. This holds also for the $s$ process in massive stars, since the role as a neutron poison is not affected by $(\gamma,n)$ reactions, even at the high temperatures during C shell burning.

\begin{acknowledgments}
We thank E.-P. Knaetsch, D. Roller and W. Seith for their help and support during the irradiations at the Van de Graaff accelerator. I.D. was supported by the Swiss National Science Foundation Grants 2024-067428.01 and 2000-105328. This work was supported by the Italian MIUR-PRIN 2006 Project "Final Phases of Stellar Evolution, Nucleosynthesis in Supernovae, AGB stars, Planetary Nebulae".
\end{acknowledgments}

%\bibliography{ca40}% Produces the bibliography via BibTeX.

\end{document}